\documentclass[12pt]{iopart}

\usepackage{graphicx}
\usepackage{amsfonts}

\usepackage{epsfig}
\usepackage{graphicx}
\usepackage{dcolumn}
\usepackage{bm}
\usepackage{xcolor}

\begin{document}
	
	\title{The optimal edge for containing the spreading of SIS model}
	
	\author{Jiajun Xian$^{1}$, Dan Yang$^{2*}$, Liming Pan$^{3,2}$, Wei Wang$^{4,2}$}
	
	\address{$^{1}$ School of Computer Science and Engineering,
		University of Electronic Science and Technology of China, Chengdu, 611731, China}
	
	\address{$^{2}$ Web Sciences Center, School of Computer Science and Engineering,
		University of Electronic Science and Technology of China, Chengdu, 611731, China}
	
	\address{$^{3}$ School of Computer Science and Technology, Nanjing Normal University, Nanjing, Jiangsu, 210023, China}
	
	\address{$^{4}$ Cybersecurity Research Institute, Sichuan University, Chengdu, 610065, China}
	\address{E-mail: *danyangsjhd@hotmail.com}

	
	\begin{abstract}    
		Numerous real--world systems, for instance, the communication platforms and transportation systems, can be abstracted into complex networks. Containing spreading dynamics (e.g., epidemic transmission and misinformation propagation) in networked systems is a hot topic in multiple fronts. Most of the previous strategies are based on the immunization of nodes. However, sometimes, these node--based strategies can be impractical. For instance, in the train transportation networks, it is dramatic to isolating train stations for flu prevention. On the contrary, temporarily suspending some connections between stations is more acceptable. Thus, we pay attention to the edge-based containing strategy.
		In this study, we develop a theoretical framework to find the optimal edge for containing the spreading of the susceptible--infected--susceptible model on complex networks. In specific, by performing a perturbation method to the discrete--Markovian--chain equations of the SIS model, we derive a formula that approximately provides the decremental outbreak size after the deactivation of a certain edge in the network. Then, we determine the optimal edge by simply choosing the one with the largest decremental outbreak size. Note that our proposed theoretical framework incorporates the information of both network structure and spreading dynamics. 
		Finally, we test the performance of our method by extensive numerical simulations. Results demonstrate that our strategy always outperforms other strategies based only on structural properties (degree or edge betweenness centrality). 
		The theoretical framework in this study can be extended to other spreading models and offers inspirations for further investigations on edge--based immunization strategies. 
		
	\end{abstract}

	
	\maketitle
	\tableofcontents
	\section{Introduction}
	The subject of containing spreading dynamics in networked systems has attracted substantial attention
	from multiple fronts, for instance, network science, statistical physics, and computer science. 
	Some common spreading dynamics, including the epidemic transmission~\cite{wang2017vaccination,wang2018characterizing} and misinformation spreading~\cite{xian2019misinformation,wang2019containing}, can influence all aspects of an individual's life and cause great impacts to the socioeconomic systems. The study of containing these spreading dynamics is of both theoretical and practical importance.
	
	Before getting into the problem of spreading containment, it is necessary to build suitable models to describe the spreading dynamics. Researchers have proposed various models for different spreading cases. For instance, the classic susceptible--infected--susceptible (SIS) model~\cite{fu2008epidemic} and the susceptible--infected--recovered (SIR) model~\cite{daley1964epidemics,daley1965stochastic}, along with many of their extensions~\cite{zuzek2015epidemic,wang2019impact,xia2019new}, have been widely applied to describe the spreading of disease or simple information (e.g., rumors). In these simple contagions, the susceptible individuals could be infected by a single contact with the infected ones. 
	When it comes to modeling some complex information spreading, such as the behavior adoption~\cite{aral2017exercise} and political information~\cite{romero2011differences,centola2007complex}, the susceptible individuals first assess the legality of the information and conduct a risk assessment. Then, they become infected with a probability that increases with the cumulative number of contacts with infected individuals. This mechanism is referred to as the social reinforcement~\cite{watts2002simple,lu2011small}. The classical threshold model and other models extended from it incorporate this mechanism for complex contagions. More spreading models with other complex mechanisms are discussed in~\cite{wang2019coevolution}.
	
	Based on different spreading models, researchers go further to develop containing strategies for the spreading dynamics. An effective containing strategy is supposed to effectively increase the spreading outbreak threshold~\cite{matsuki2019intervention} or decrease the outbreak size~\cite{xian2019containing}. 
	Many target containing strategies have been proposed, for instance, immunizing a fraction of nodes according to the centrality indexes of them, like their degree, betweenness, closeness, PageRank, and eigenvector centrality~\cite{wang2016statistical,holme2002attack,cohen2003efficient,chung2009distributing,prakash2013fractional,lu2016vital,zhang2013impacts}. However, in the real--life, the identification of centrality--defined individuals sometimes can be time--consuming. Thus, researchers come up with strategies that do not rely on any centrality indexes of the nodes, such as the acquaintance immunization strategy~\cite{wang2015immunity,cohen2003efficient,liu2014controlling,wang2016statistical,wang2015immunity}, which is more suitable for practical applications. Besides, many containing strategies inspired by the methods from optimization and control~\cite{morone2015influence,preciado2014optimal,preciado2014convex,nowzari2016analysis,wang2016suppressing,chen2018suppressing,zhang2014suppression} have been proposed by researchers as well. There are also some specific containing strategies for networks in different categories, for instance, the multiple networks~\cite{zhao2014immunization}, adaptive networks~\cite{ogura2017optimal}, and temporal networks~\cite{starnini2013immunization}.
	
	All the strategies mentioned above are based on the immunization of nodes. However, these strategies are difficult to be put into practice sometimes. For instance, it is dramatic to isolate a fraction of train stations in the whole country for preventing the spreading of flu on the train transportation network, but it is more acceptable to suspend some connections in the network. Thus, the study of edge--based containing strategies should be emphasized. Some researchers have proposed strategies of deactivating edges selected by the properties of the adjacent nodes or the edges themselves~\cite{schneider2011suppressing,van2011decreasing}. Besides, some strategies incorporate both the structural characteristics of the network and parameters of the spreading process~\cite{matamalas2018effective}.
	
	This study focus on the subject of determining the optimal edge for containing the spreading of the SIS model on complex networks. The theoretical framework we developed can find out the optimal or near--optimal edge for the spreading containment of the SIS model. By developing a perturbation method to the discrete--Markovian--chain equations of the SIS model, we obtain a formula that approximately provides the decremental outbreak size after deactivating a certain edge in the network. Then, we determine the optimal edge by simply selecting the one with the largest decremental outbreak size. It is worth mentioning that the information of both network structure and spreading dynamics is considered in our theoretical framework. 
	
	The paper is organized as follows. Sec. 2 provides the model description. The detailed theoretical analysis is presented in Sec. 3. Then, we present the numerical simulations in Sec. 4. Finally, we provide a conclusion in Sec. 5.
	
	\section{Model description} \label{sec:model}
	
	In this study, we consider the classic SIS model on a complex network $G$ of $N$ nodes and $M$ edges. The SIS model is extensively applied to describe the spreading of simple information or disease.
	Each node in this model can be in two different states, that is, the susceptible state (S) and the infected state (I).
	Initially, we select a small fraction of nodes to be the infected seeds, keeping the others in the S state.
	Then, for every time step, each infected node tries to infect neighbors in the S state with probability $\lambda$.
	Afterward, all the nodes in the I state return to the S state with probability $\gamma$. 
	Without loss of generality, we set $\gamma=0.5$ in this study.
	Eventually, the dynamic system will reach the steady--state on network $G$, where the fraction of nodes in the I state fluctuates around a certain value $\rho$, that is, the outbreak size. 
	
	Let $A$ be the adjacency matrix of the network $G$. 
	Thus, $A$ should be a square $N \times N$ matrix such that its element $A_{ij}=1$ when there is an edge between node $i$ and $j$, and $A_{ij}=0$ when there is no edge.
	Previous studies~\cite{de2017disease,gomez2010discrete} have demonstrated that the spreading outbreaks when the effective transmission probability $\beta=\lambda/\gamma$ is larger than the reciprocal of the leading eigenvalue $\omega$ of $A$. That is to say, if $\lambda_c/\gamma=1/\omega$, then the spreading will break out only when $\lambda$ is larger than the critical value $\lambda_c=\gamma/\omega$. Otherwise, if $\lambda<\lambda_c$, then no outbreaks will be observed, and no containing process is needed. Therefore, we focus on the case when $\lambda>\lambda_c$ in this study.
	
	After deactivating a specific edge $(i_0,j_0)$ in the networks $G$, we get a new network $G^{'}$. And the adjacency matrix $A^{'}$ of $G^{'}$ should be
	\begin{equation}\label{eq:aaa}
	A^{'}=A-\dot{ A},
	\end{equation}
	where the element $\dot{ A}_{ij}=1$ only when $(i,j) \in \{(i_0,j_{0}),(j_{0},i_0)\}$. Denote $\rho^{'}$ as the outbreak size on the new network $G^{'}$.
	We aim to find the optimal edge, which, if deactivated, can maximize the decremental outbreak size $\dot \rho=\rho-\rho^{'}$.

	\section{Theoretical analysis} \label{sec:theory}
	In this section, we first present the Discrete--Markovian--chain (DMC) approach~\cite{gomez2010discrete,pan2019optimal} for the SIS model on the network $G$. Then, using a perturbation method for the DMC, we derive a formula that approximately provides the decremental outbreak size after deactivating an edge in the network $G$. Finally, using the formula, we study the problem of determining the optimal edge, which, if deactivated, can maximize the decremental outbreak size.
	\subsection{The Discrete--Markovian--chain approach for the SIS model}\label{DMC}
	
	In this subsection, we adopt the discrete--Markovian--chain (DMC) approach to study the SIS model on the network $G$. The DMC approach can accurately predict the phase diagram for contact--based spreading dynamics in complex networks and overcomes the computational cost of Monte Carlo simulations~\cite{gomez2010discrete}. 
	
	To begin with, we define a set of discrete--time equations for the probability of individual nodes to be infected.
	Denote $I_i(t)$ as the probability that node $i$ is in the I state at time $t$. Then, the node $i$ is in the S state at time $t$ with probability $S_i(t) = 1-I_i(t)$.
	If $i$ is in the I state at $t+1$, then either it was in the I state at $t$ and has not recovered, or it was in the S state at $t$ and has been infected by its infected neighbors. Thus, the evolution of $I_i(t)$ is
	\begin{equation}\label{eq:pit}
	I_i(t+1)=(1-\gamma)I_i(t)+[1-\Theta_i(t)]S_i(t),
	\end{equation}
	where $1-\Theta_i(t)$ is the probability that node $i$ gets infected at time $t$, and
	\begin{equation}\label{eq:qit}
	\Theta_i(t)=\prod_{j=1}^{N}[1-\lambda A_{ij}I_j(t)].
	\end{equation}
	When the dynamic system reaches the steady state, we have $I_i(t)=I_i(t+1)=
	\widetilde I_i$, $S_i(t)=S_i(t+1)=
	\widetilde S_i$, and $\Theta_i(t)=\Theta_i(t+1)=\widetilde \Theta_i$. Taking all the nodes into consideration, the Eqs.~(\ref{eq:pit}) and (\ref{eq:qit}) can be written in terms of vectors in  the steady state as
	\begin{equation}\label{eq:fixEq1}
	\widetilde I=(1-\gamma)\widetilde I+(1- \widetilde \Theta)\circ \widetilde S,
	\end{equation}
	and
	\begin{equation}\label{eq:fixEq2}
	\widetilde \Theta_i=\prod_{j=1}^N(1-\lambda A_{ij} \widetilde I_j),
	\end{equation}
	where $\widetilde I$ and $\widetilde \Theta$ are vectors of length $N$ with entries $\widetilde I=(\widetilde I_1,\cdots,\widetilde I_N)^{\mathrm{T}}$ and
	$\widetilde \Theta=(\widetilde \Theta_1,\cdots,\widetilde \Theta_N)^{\mathrm{T}}$, respectively, and $\circ$ denotes component--wise vector product.
	Combing the Eqs.~(\ref{eq:fixEq1}) and (\ref{eq:fixEq2}), we obtain the expected outbreak size of the spreading on networks $G$ as follows:
	\begin{equation}
	\rho= N^{-1}\mathbf{1}^{\mathrm{T}}\widetilde I.
	\end{equation}

	\subsection{Determining the optimal edge for containing the spreading}\label{PM}
	
	For convenience, we denote the new network we get after deactivating the specific edge $l$ in the original network by $G^{'}_{l}$. Besides, the decremental outbreak size $\dot \rho$ of the SIS model on network $G$ after deactivating the specific edge $l$ is denoted by $\dot \rho_{l}$. 
	The spreading outbreak size $\rho_{l}$ on the new network $G_{l}$ depends on the position of edge $l$.
	In this subsection, we introduce a perturbation method to obtain an approximate estimate of the decremental outbreak size $\dot \rho_{l}=\rho-\rho_{l}$ after the deactivation of edge $l$. Then, we use the approximate estimate to determine the optimal edge, which, if deactivated, can maximize the decremental outbreak size $\dot \rho$.

	Inferring from Eq.~(\ref{eq:aaa}), we get the DMC equations of network $G_{l}$ as follows:
	\begin{equation}\label{eq:pit2}
	I_i(t+1)=(1-\gamma)I_i(t)+[1-\Theta_i(t)][1-I_i(t)],
	\end{equation}
	and 
	\begin{equation}\label{eq:qit2}
	\Theta_i(t)=\prod_{j=1}^{N}[1-\lambda (A_{ij}-\dot{ A}_{ij})I_j(t)].
	\end{equation}
	On the consideration that the fixed point $I(\infty)=[I_1(\infty),\cdots,I_N(\infty)]^{\mathrm{T}}$ of $G_{l}$ will stays close to the fixed point $\widetilde I$ of $G$ since only one edge has been deactivated, we iterate Eqs.~(\ref{eq:pit2}) and (\ref{eq:qit2}) with initial condition $I(0)=\widetilde I$. Then, we employ the decomposition of $I(t)=\widetilde I+\dot I(t)$ and $\Theta(t)=\widetilde \Theta+\dot \Theta(t)$. According to \ref{It}, we can obtain the iteration formula of $\dot I(t)$ as 
	\begin{eqnarray}
	\dot{ I}(t+1)&=&(\widetilde \Theta-\gamma)\dot{ I}(t)
	+(1-\widetilde I)\circ\lambda \widetilde \Theta\circ \left(A-\dot{ A}\right) \Psi \dot{ I}(t) \nonumber \\
	& &+(1-\widetilde I)\circ \widetilde \Theta\circ \dot{ A}\log(1-\lambda \widetilde I),
	\end{eqnarray}
	where $\Psi$ is the $N\times N$ diagonal matrix  with entries $\Psi_{ij} =1/(1-\lambda \widetilde I_j)$ for $i = j$ and $\Psi_{ij} =0$ for $i\neq j$. 
	This equation can be written in terms of matrix multiplication as follows:
	\begin{equation}\label{eq:pertuRelationMatrix}
	\dot{ I}(t+1)=\Xi\dot{ I}(t)+\xi,
	\end{equation}
	where 
	\begin{equation}\label{A}
	\Xi=\lambda \mathrm{diag}(\widetilde \Theta-\widetilde I\circ \widetilde \Theta)(A-\dot{ A})\Psi+\mathrm{diag}(\widetilde \Theta-\gamma)
	\end{equation}
	and
	\begin{equation}\label{B}
	\xi=(1-\widetilde I)\circ \widetilde \Theta\circ \dot{ A}\log(1-\lambda \widetilde I).
	\end{equation}
	Here, $\mathrm{diag}\left(\cdot\right)$ denotes the diagonal matrix with the elements of the input vector as diagonal entries. Thus, the stationary solution $\dot I(\infty)$ of the perturbed system satisfies
	\begin{equation}
	\dot I(\infty)=\Xi\dot I(\infty)+\xi,
	\end{equation}
	or in the closed form
	\begin{equation}\label{eq:dprelation}
	\dot I(\infty)=\left(\mathbb{I}-\Xi\right)^{-1}\xi.
	\end{equation}
	Thus, an explicit relation between the deactivated edge $l$ and the decremental outbreak size can be found as
	\begin{equation}\label{eq:dp}
	\dot \rho_{l}=-N^{-1}\mathbf{1}^{\mathrm{T}}\dot I(\infty)=-N^{-1}\mathbf{1}^{\mathrm{T}}\left(\mathbb{I}-\Xi\right)^{-1}\xi.
	\end{equation}
	
	In order to solve Eq.~(\ref{eq:dp}), we decompose $\Xi$ as $\Xi=\Xi^0+\dot \Xi$, where
	\begin{equation}\label{Xi0}
	\Xi^0 =\lambda \mathrm{diag}(\widetilde \Theta-\widetilde I\circ \widetilde \Theta)A \Psi+\mathrm{diag}(\widetilde \Theta-\gamma)
	\end{equation}
	depends only on $A$, and
	\begin{equation}
	\dot \Xi =-\lambda \mathrm{diag}(\widetilde \Theta-\widetilde I\circ \widetilde \Theta)\dot A \Psi
	\end{equation}
	depends only on $\dot A$.
	For any edge $l=(i,j)$, the matrix $\dot A$ can be written as sum of two outer products
	\begin{equation}
	\dot A=uv^{\mathrm{T}}+vu^{\mathrm{T}},
	\end{equation}
	where $u$, $ v$ are vectors of length $N$ with $u_k=\delta_{k,i}$ and $v_{k}=\delta_{k,j}$ for $1\leq k\leq N$.
	Define short notations as
	\begin{equation}\label{eij}
	\varepsilon_{ij}\mathop{:}=-\lambda \left(\widetilde \Theta_i-\widetilde I_i \widetilde \Theta_i\right)\left(1-\lambda \widetilde I_j\right)^{-1},
	\end{equation}
	then it's easy to check that
	\begin{equation}
	\dot \Xi= \varepsilon_{ij} uv^{\mathrm{T}}+\varepsilon_{ji} vu^{\mathrm{T}}.
	\end{equation}
	The Sherman-Morrison formula says that
	\begin{equation}\label{eq:Cmatrix}
	(\mathbb{I}-\Xi)^{-1}=\left(\mathbb{I}-\Xi^0-\varepsilon_{ij} uv^{\mathrm{T}}-\varepsilon_{ji} vu^{\mathrm{T}}\right)^{-1}
	=X+\frac{\varepsilon_{ij}X u v^{\mathrm{T}}X}{1-\varepsilon_{ij}X_{ji}},
	\end{equation}
	where
	\begin{equation}
	X=\left(\mathbb{I}-\Xi^0-\varepsilon_{ij} uv^{\mathrm{T}}\right)^{-1}.
	\end{equation}
	Apply the Sherman-Morrison formula again gives
	\begin{equation}\label{xy}
	X=Y+\frac{\varepsilon_{ji}Y v u^{\mathrm{T}}Y}{1-\varepsilon_{ji}Y_{ij}},
	\end{equation}
	where
	\begin{equation}\label{Y}
	Y=\left(\mathbb{I}-\Xi^0\right)^{-1}.
	\end{equation}
	Again define short notations for convenience as follows,
	\begin{equation}\label{cij}
	c_{ij}=\left(\widetilde \Theta_i-\widetilde I_i \widetilde \Theta_i\right)\log\left(1-\lambda \widetilde I_j\right).
	\end{equation}
	Then $\xi$ can be checked satisfying
	\begin{equation}\label{eq:xi}
	\xi= c_{ij} u+c_{ji} v.
	\end{equation}
	By substituting Eq.~(\ref{eq:Cmatrix}) and Eq.~(\ref{eq:xi}) into Eq.~(\ref{eq:dp}), we can get
	\begin{eqnarray}\label{eq:np1}
	\dot \rho_{l}
	&=&c_{ij} \mathbf{1}^{\mathrm{T}}Xu+c_{ji} \mathbf{1}^{\mathrm{T}}Xv+\frac{\varepsilon_{ij}c_{ij} X_{ji}\mathbf{1}^{\mathrm{T}}X u}{1-\varepsilon_{ij}X_{ji}}\nonumber \\
	& &+\frac{\varepsilon_{ij}c_{ji} X_{jj}\mathbf{1}^{\mathrm{T}}X u}{1-\varepsilon_{ij}X_{ji}}.
	\end{eqnarray}
	According to Eq.~(\ref{xy}), we carefully expand Eq.~(\ref{eq:np1}) and get
	\begin{eqnarray}\label{eq:dpformula}
	\dot \rho_{l}&=&-N^{-1}[\frac{\left(c_{ij}-c_{ij}\varepsilon_{ji}Y_{ij}+c_{ji} \varepsilon_{ij}Y_{jj}\right)\mathbf{1}^{\mathrm{T}}Yu}{\left(1-\varepsilon_{ij}Y_{ji}\right)\left(1-\varepsilon_{ji}Y_{ij}\right)-\varepsilon_{ij}\varepsilon_{ji}Y_{ii} Y_{jj}}\nonumber \\
	& &+\frac{\left(c_{ji} - c_{ji}\varepsilon_{ij} Y_{ji} +c_{ij}\varepsilon_{ji} Y_{ii}\right)\mathbf{1}^{\mathrm{T}}Yv}{\left(1-\varepsilon_{ij}Y_{ji}\right)\left(1-\varepsilon_{ji}Y_{ij}\right)-\varepsilon_{ij}\varepsilon_{ji}Y_{ii} Y_{jj}}].
	\end{eqnarray}
	Eq.~(\ref{eq:dpformula}) approximately provides the decremental outbreak size after deactivating the edge $l=(i,j)$ in the network $G$.
	We can use the formula to determine the optimal edge for containing the spreading of the SIS model by simply selecting the edge with the highest $\dot \rho_{l}$. 
	As one can see, we obtain Eq.~(\ref{eq:dpformula}) through complicated derivations.
	Look into the right-hand of Eq.~(\ref{eq:dpformula}), we can observe that it incorporates both the information of networks structure (i.e., the adjacency matrix $A$) and spreading dynamics (i.e., $\lambda$, $\gamma$, $\widetilde{I}$ and $\widetilde{\Theta}$). Sec.~\ref{sec:simulation}
	will show that Eq.~(\ref{eq:dpformula}) gives good predictions of the decremental outbreak size.

	\section{Simulation results}\label{sec:simulation}

	For convenience, we refer to the containing strategy of deactivating the optimal edge $L$ selected by Eq.~(\ref{eq:dpformula}) as the perturbation--based--strategy (PBS).
	In this section, extensive numerical simulations are performed to verify the containing performance of the PBS. 
	Both synthetic and real--world networks are considered in our simulations. Note that the DMC approach can predict the results of the Monte Carlo simulations accurately and have a lower computational cost; thus, we conduct our numerical simulations based on the DMC approach instead of the Monte Carlo method~\cite{gomez2010discrete}.

	According to Sec. \ref{sec:theory}, the proposed PBS in this study incorporates the information of both network structure and spreading dynamics. To better understand the importance of dynamic information in the PBS, we employ two contrast strategies that only consider the structure of networks. 
	Denote $f^b$ as the edge betweenness centrality.  
	Besides, the edge betweenness of the specific edge $l=(i,j)$ is denoted by $f^b_{l}$.  
	The first contrast strategy is to deactivate the edge with the highest $f^b$. We refer to this strategy as betweenness--centrality--strategy (BCS) and the specific edge selected by the BCS as $L^{B}$. Similarly, the second contrast strategy is based on the degree $k$ of nodes; thus, it can be referred to as degree--based--strategy (DBS). Specifically, for the DBS, we select the edge with the highest degree product $f^d$. For the specific edge $l=(i,j)$, the degree product should be $f^d_{l}=k_{i}k_{j}$. The edge selected by the DBS is denoted by $L^{D}$.

\begin{figure}
	\centering
	\includegraphics[width=0.9\textwidth]{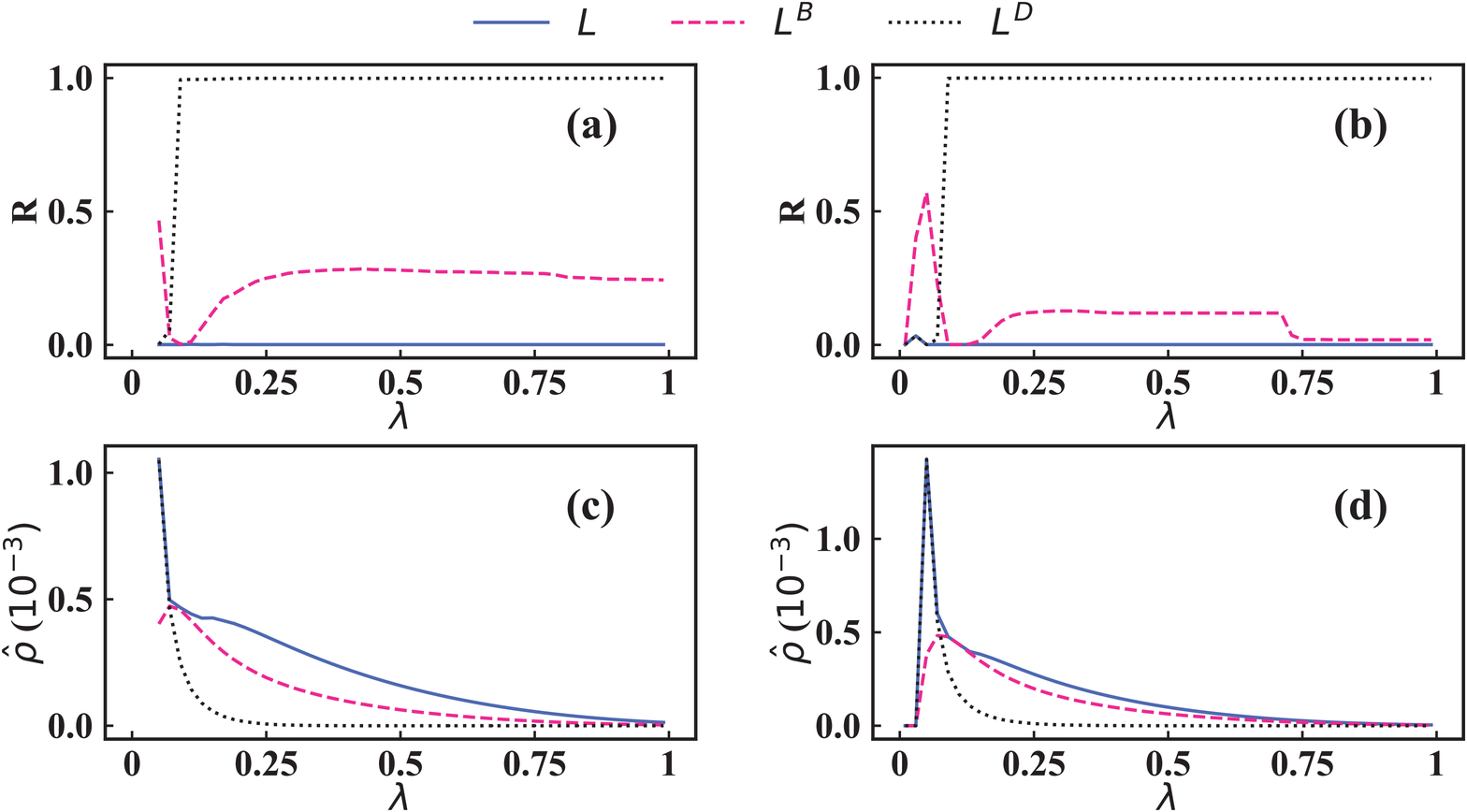}
	\caption{(Color online) Performance of different strategies versus $\lambda$  on synthetic networks. The normalized numerical rank $R$ of the optimal edges $L$ (blue solid line), $L^{B}$ (red dashed line), and $L^{D}$ (black dotted line) on the SF networks with (a) $\alpha_{1}=2.3$, and (b) $\alpha_{2}=3.0$. 
	The decremental outbreak size $\hat \rho $ obtained by simulations after deactivating the optimal edge $L$ (blue solid line), $L^{B}$ (red dashed line), and $L^{D}$ (black dotted line) on the SF networks with (c) $\alpha_{1}=2.3$, and (d) $\alpha_{1}=3.0$. 
	More information about the two synthetic networks can be found in Tab.~\ref{tab:networks}.}\label{numrical_rank}
\end{figure}

	First, we perform the containing strategies on two synthetic networks $G^1$ and $G^2$. Both of them are scale--free (SF) networks with degree distribution $p(k)\sim k^{-\alpha}$, where $\alpha$ denotes the degree exponent. We set $\alpha_{1}=2.3$ and $\alpha_{2}=3.0$ as the degree exponent of network $G^1$ and $G^2$, respectively.
	One can find more information about the two networks in Tab.~\ref{tab:networks}. 
	Denote $\hat \rho$ as the decremental outbreak size obtained by simulations after deactivating the selected edge. 
	Then, we rank the edges according to the values of $\hat \rho$. We refer to this kind of edge rank as numerical rank $r$ in the rest of the paper.  
	To compare the performance of the strategies, we are particularly concerned about the optimal edges $L$, $L^{B}$, and $L^{D}$ selected by PBS, BCS, and DBS, respectively. 
	We compute the normalized numerical rank $R=r/M$ of edges $L$, $L^{B}$ and $L^{D}$. The smaller the $R$, the better the performance.
	As shown in the Figs.~\ref{numrical_rank} (a) and (b), the DBS performs well on both synthetic networks when $\lambda$ near the critical value $\lambda_c$, but the performance fails quickly when $\lambda$ becomes large. The BCS performs well only for several values of $\lambda$. However, the PBS performs well on both networks for all the values of $\lambda$.
	Figs.~\ref{numrical_rank} (c) and (d) show the corresponding $\hat \rho$ of edges $L$, $L^{B}$ and $L^{D}$ on the synthetic networks $G^1$ and $G^2$, respectively. The larger the $\hat \rho$, the better the performance. By comparing the $\hat \rho$ of $L$, $L^{B}$ and $L^{D}$, we can draw the same conclusion as that demonstrated from Figs.~\ref{numrical_rank} (a) and (b).  

\begin{figure}
		\centering
		\includegraphics[width=0.9\textwidth]{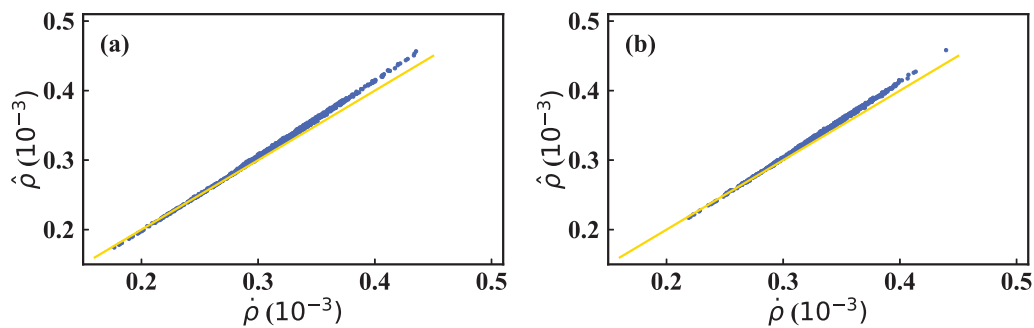}
		\caption{(Color online) The correlation between decremental outbreak size $\dot \rho$ and $\hat \rho$. The correlation between decremental outbreak size $\dot \rho$ numerically computed by Eq.~(\ref{eq:dpformula}) and the decremental outbreak size $\hat \rho$ obtained by simulations on the SF networks with (a) $\alpha_{1}=2.3$, and (b) $\alpha_{2}=3.0$. The dynamical parameters are set to be $\lambda=0.1$ and $\gamma=0.5$. The yellow solid lines in the plots represent the function $\hat \rho=\dot \rho$.
	 }\label{yeqx}
	\end{figure}

\begin{figure}
		\centering
		\includegraphics[width=0.9\textwidth]{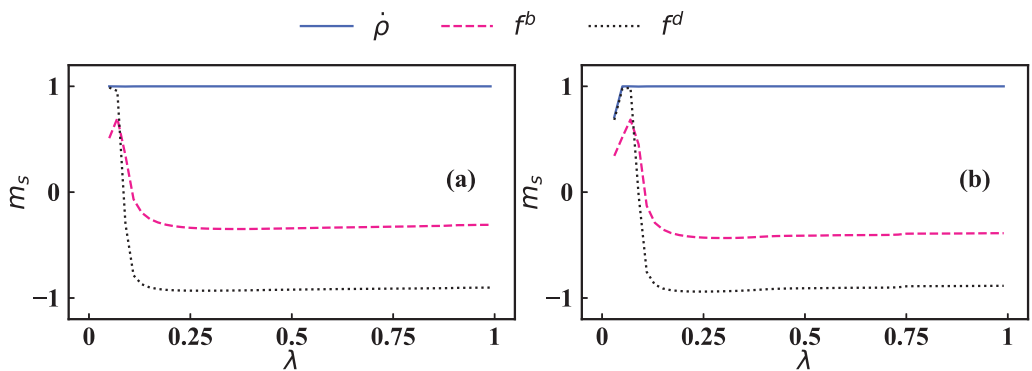}
		\caption{(Color online) The correlations between the approximate edge ranks and the numerical edge ranks. The Spearman’s rank correlation coefficient $m_s$ between the edge ranks scored by Eq.~(\ref{eq:dpformula}) (blue solid line) and the numerical ranks versus $\lambda$ on the SF networks with (a) $\alpha_{1}=2.3$, and (b) $\alpha_{2}=3.0$. The Spearman’s rank correlation coefficient between the edge ranks scored by the edge betweenness centrality $f^b$ and the numerical ranks are denoted by red dashed lines. Black dotted lines denote the Spearman’s rank correlation coefficient between the edge ranks scored by degree product $f^d$ and the numerical ranks.
		More information about the two synthetic networks can be found in Tab.~\ref{tab:networks}. }\label{ms}
	\end{figure}

	Second, we investigate the overall correlations between the edge ranks scored by Eq.~(\ref{eq:dpformula}) and the numerical ranks. To begin with, we compare the decremental outbreak size $\dot \rho$ numerically computed by Eq.~(\ref{eq:dpformula}) with the decremental outbreak size $\hat \rho$ obtained by simulations.
	Setting $\lambda=0.1$, the results of $\hat \rho$ versus $\dot \rho$ on the synthetic networks $G^1$ and $G^2$ are shown in the Figs.~\ref{yeqx} (a) and (b), respectively. The results demonstrate that the values of $\hat \rho$ and $\dot \rho$ are almost linearly correlated; that is to say, Eq.~(\ref{eq:dpformula}) can predict the decremental outbreak size well. 
	To better understand the rank correlations for all the values of $\lambda$, we employ the Spearman rank correlation coefficient~\cite{lee2012correlated,cho2010correlated} to quantify the mentioned correlations. The Spearman rank correlation coefficient is defined as 
	\begin{equation}\label{eq:xiSIR}
	m_s=1-6\frac{\sum_{l=1}^M(\hat r_{l}-\dot r_{l})^{2}｝}{M(M^2-1)},
	\end{equation}
	where $\hat r_{l}$ and $\dot r_{l}$ are the ranks of edge $l$ scored by $\hat \rho$ and $\dot \rho$, respectively. Figs.~\ref{ms} (a) and (b) shows the results of $m_s$ versus $\lambda$ on the synthetic networks $G^1$ and $G^2$, respectively. It can be seen that the value of $m_s$ keeps close to 1 for all the values of $\lambda$ on both networks. That is to say, the edge ranks predicted by the Eq.~(\ref{eq:dpformula}) and the numerical ranks are strongly correlated for all the values of $\lambda$. Similarly, we also compute the Spearman rank correlation coefficient between the edge ranks scored by $f^b$ and the numerical ranks, along with the Spearman rank correlation coefficient between the edge ranks scored by $f^d$ and the numerical ranks. As shown in Figs.~\ref{ms} (a) and (b), the edge ranks scored by $f^b$ or $f^d$ is positively correlated with the numerical ranks only when $\lambda$ is near the $\lambda_c$. When $\lambda$ becomes large, their correlations become negative. The results in Figs.~\ref{ms} (a) and (b) demonstrate that the Eq.~(\ref{eq:dpformula}) is sufficient to predict the numerical rank of edges, but the $f^b$ or $f^d$ is far from sufficient.
	
	\begin{table}
		\centering
		\caption{Basic statistics of the two synthetic networks and six real--world networks employed in this study: the number of
			nodes $N$, the number of edges $M$, the average degree $\left\langle k\right\rangle $, and the theoretical spreading threshold $\lambda_c$.}
		\begin{tabular}{lllllll}
			\br
			Name                    &$N$ & $M$ &  $\left\langle k\right\rangle $  & $\lambda_c$     \\ \mr
			SF2.3                   & 200  & 1000      & 10                                            & 0.076 \\
			SF3.0                  & 200  & 1000     & 10                                        & 0.083 \\
			Residence hall          & 217  & 1839     & 16.949                                           & 0.046 \\
			Hamsterster friendships & 1788 & 12476   & 13.955                                           & 0.022 \\
			Jazz musicians          & 198  & 2742    & 27.697                                          & 0.025 \\
			Facebook (NIPS)         & 2888 & 2981    & 2.0644                                         & 0.036 \\
			Physicians              & 117  & 465     & 7.95                                              & 0.099 \\
			Air traffic control     & 1226 & 2408     & 3.928                                             & 0.109 \\ \br
		\end{tabular}
		\label{tab:networks}
	\end{table}

	\begin{figure}
	\centering
	\includegraphics[width=0.9\textwidth]{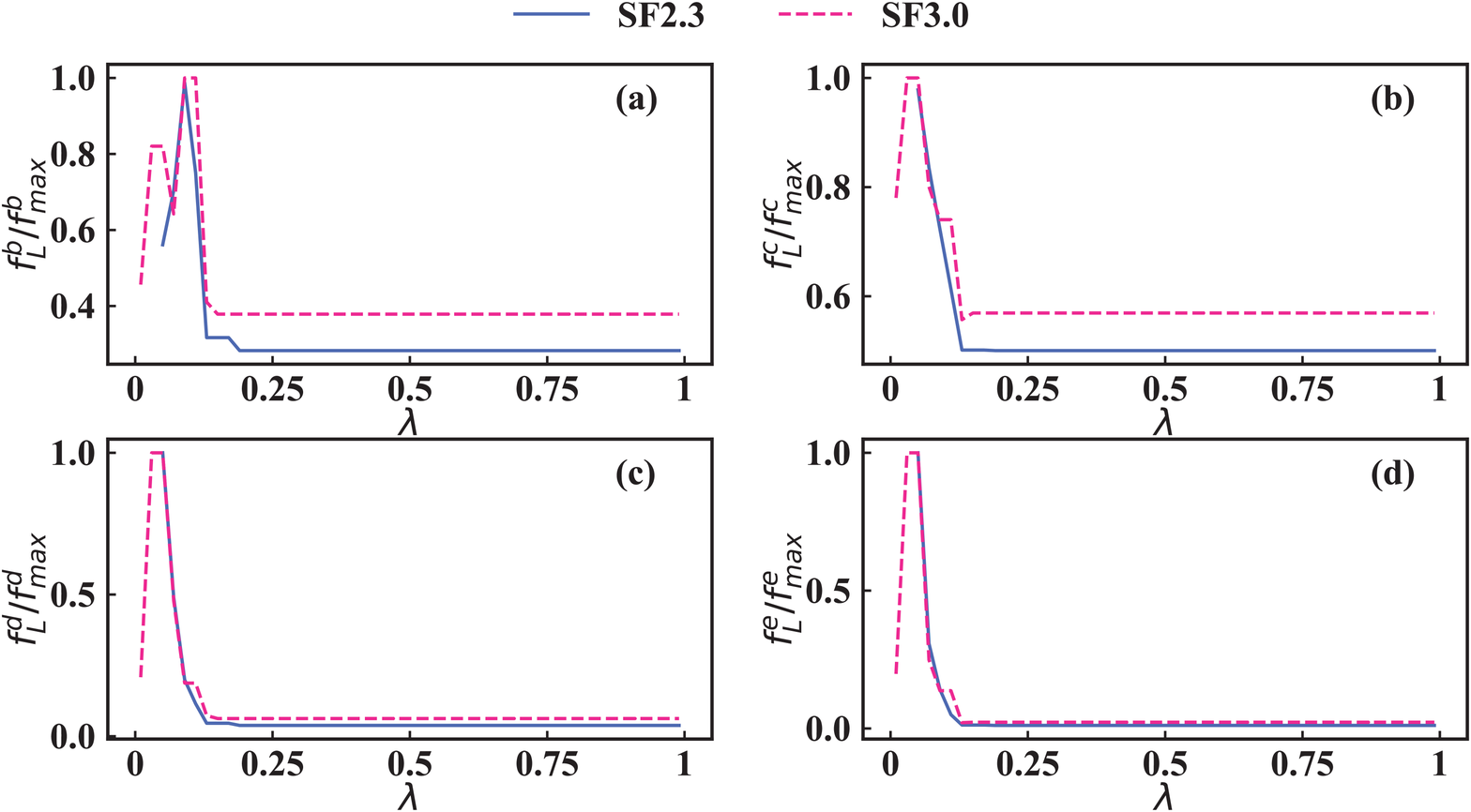}
	\caption{(Color online) Normalized structural properties of the optimal edge $L$ selected by Eq.~(\ref{eq:dpformula}). (a) $f^b_{L}/f^b_{max}$ versus $\lambda$, that is, the normalized edge betweenness centrality of $L$. (b) $f^c_{L}/f^c_{max}$ versus $\lambda$, that is, the normalized product of the closeness centrality of nodes at the two ends of $L$. (c) $f^d_{L}/f^d_{max}$ versus $\lambda$, that is, the normalized product of the degree of nodes at the two ends of $L$. (d) $f^e_{L}/f^e_{max}$ versus $\lambda$, that is, the normalized product of the eigenvector centrality of nodes at the two ends of $L$. Blue solid lines (red dashed lines) denotes the corresponding results of the SF network with $\alpha_{1}=2.3$ ($\alpha_{2}=3.0$). Some structural properties about the two synthetic networks can be found in Tab.~\ref{tab:networks}.    }\label{statistics}
\end{figure}
	
	We now go further to investigate the structural properties of the optimal edge $L$ selected by Eq.~(\ref{eq:dpformula}). The normalized structural statistics $f^b_{L}/f^b_{max}$, $f^c_{L}/f^c_{max}$, $f^d_{L}/f^d_{max}$ and $f^e_{L}/f^e_{max}$ versus $\lambda$ are shown in Figs.~\ref{statistics} (a)--(d), respectively, where $f^c_{L}$ ($f^e_{L}$) denotes the product of the closeness centrality (eigenvector centrality) of the nodes at the two ends of edge $L$. Note that $f^{x}_{max}$ is the maximum value in $\{f^{x}_{l}\}$, where $1\leq l\leq M$ and $x \in \{b,c,d,e\}$. As demonstrated in Fig.~\ref{statistics}, when the value of $\lambda$ is small, the optimal edge $L$ has large $f^b_{L}/f^b_{max}$, $f^c_{L}/f^c_{max}$, $f^d_{L}/f^d_{max}$, and $f^e_{L}/f^e_{max}$. That is to say, when $\lambda$ is slightly above the critical point $\lambda_c$, the optimal edges should be those have high edge betweenness centrality and connect nodes with high closeness, degree, and eigenvector centrality. This can be explained by the fact that the outbreak size is small near the critical point, and the edge between nodes with high centrality can help to keep the cluster of nodes in I state. Thus, deactivating the edge with high centrality can well contain the spreading when $\lambda$ is small.
	However, when the $\lambda$ becomes large, the optimal edge $L$ will have low $f^d_{L}/f^d_{max}$, $f^e_{L}/f^e_{max}$ and middle $f^b_{L}/f^b_{max}$, $f^c_{L}/f^c_{max}$. This is because nodes with high centrality will have a high probability of being infected when $\lambda$ is large; thus, deactivating the edge between nodes with high centrality becomes unnecessary.

\begin{figure}
	\centering
	\includegraphics[width=0.9\textwidth]{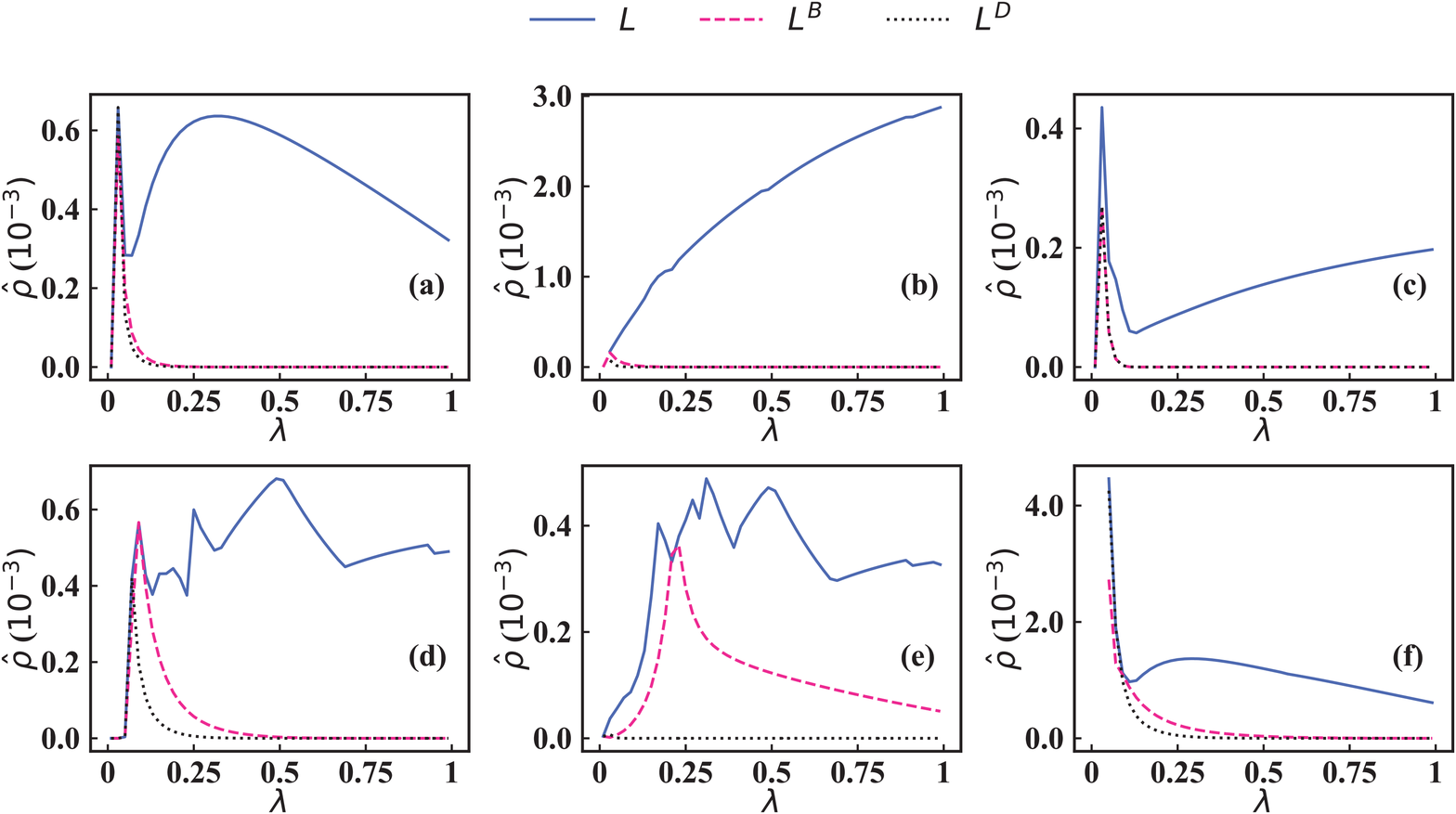}
	\caption{(Color online) Performance of different strategies on real--world networks. The decremental outbreak size $\hat \rho $ obtained by simulations after deactivating the optimal edge $L$ (blue solid line), $L^{B}$ (red dashed line), and $L^{D}$ (black dotted line) on the real--world networks (a) Residence hall, (b) Jazz musicians, (c) Facebook (NIPS), (d) Air traffic control, (e) Hamsterster friendships, and (f) Physicians.
	Detailed information about the six real--word networks can be found in Tab.~\ref{tab:networks}.  }\label{real-network}
	\end{figure}

	Finally, we test the performance of the strategies on six real--world networks: (a) Residence hall~\cite{konect:freeman1998}, (b) Jazz musicians~\cite{konect:arenas-jazz}, (c) Facebook (NIPS)~\cite{konect:McAuley2012}, (d) Air traffic control~\cite{konect:faa}, (e) Hamsterster friendships~\cite{konect:2017:petster-friendships-hamster}, and (f) Physicians~\cite{konect:coleman1957}. Tab.~\ref{tab:networks} provides some basic statistics of these networks. More detailed information of these networks can be found in~\cite{real-network}, where they are downloaded from.
	Figs.~\ref{real-network} (a)--(i) show the decremental outbreak size $\hat \rho$ obtained by simulations after deactivating the optimal edges $L$, $L^B$ or $L^D$ on the six real-world networks for different transmission probability $\lambda$. The results demonstrate that the PBS outperforms the two contrast strategies on all the six real--networks and for all the values of $\lambda$ studied.

	\section{Conclusions}\label{sec:conclusion}
	
	Containing spreading dynamics (e.g., epidemic transmission and misinformation propagation) in the networked systems (e.g., transportation systems and communication platforms) is of both theoretical and practical importance. In this study, we developed a theoretical framework to determine the optimal edge for containing the spreading of the SIS model on complex networks.
	
	To be specific, we performed a perturbation method to the DMC equations of the SIS model and obtained a formula that provides an approximate value of the decremental outbreak size after deactivating a certain edge in the network. Afterward, we determined the optimal edge by selecting the one with the largest decremental outbreak size. It is worth mentioning that the formula we obtained incorporates the information of both network structure and spreading dynamics. Extensive numerical simulations on both synthetic networks and real--world networks demonstrated that our strategy performs well for all the values of $\lambda$ and outperforms those strategies based only on structure statistics (degree or edge betweenness centrality).
	
	Previous strategies of containing spreading dynamics on complex networks are mostly based on node immunization, which can be socially and politically difficult in practice sometimes. The theoretical framework developed in this study offers inspirations for investigations on edge--based immunization strategies, which could be more practical for some specific real situations. Our theoretical framework could also be extended to other spreading models.

	\section*{Acknowledgements}
	
	This work was partially supported by the China Postdoctoral Science Foundation (Grant No.~2018M631073), China Postdoctoral Science Special Foundation (Grant No.~2019T120829), National Natural Science Foundation of China (Grant Nos.~61903266 and 61603074), and Fundamental Research Funds for the Central Universities.
	
	\appendix
	\section{}\label{It}
	
	\begin{center}
		\textbf{The iteration formula of $\dot I(t)$}
	\end{center}
	
	This appendix shows the detailed steps of obtaining the iteration formula of $\dot I(t)$. Based on the decomposition of $I(t)=\widetilde I+\dot I(t)$ and $\Theta(t)=\widetilde \Theta+\dot \Theta(t)$, we get
	\begin{equation} \label{add_p}
	\widetilde I+\dot I(t+1)=(1-\gamma)(\widetilde I+\dot I(t))+(1-\widetilde I-\dot I(t))\circ (1-\widetilde \Theta-\dot \Theta(t)),
	\end{equation}
	where $\dot I(t)$ and $\dot \Theta(t)$ are assumed small. Ignoring the second-order term $\dot I(t) \circ \dot \Theta(t)$, we can get 
	\begin{equation}\label{eq:fixPertuEq1}
	\dot I(t+1)=(\widetilde \Theta-\gamma)\dot I(t)-(1-\widetilde I)\circ \dot \Theta(t)
	\end{equation}
	by expanding Eq.~(\ref{add_p}) and substituting $\widetilde I=(1-\gamma)\widetilde I+(1- \widetilde \Theta)\circ \widetilde S$.
	Similarly, $\Theta(t)$ becomes
	\begin{equation}\label{qi_f}
	\widetilde \Theta_i+\dot \Theta_i(t)=\prod_{j=1}^N\{1-\lambda (A_{ij}-\dot A_{ij})[\widetilde I_{j}+\dot I_{j}(t)]\}.
	\end{equation}
	We notice the following equation holds
	\begin{equation}\nonumber
	[1-\lambda (A_{ij}-\dot A_{ij})][\widetilde I_j+\dot I_j(t)]
	=\frac{1-\lambda A_{ij}[\widetilde I_j+\dot I_j(t)]}{1-\lambda \dot A_{ij}[\widetilde I_j+\dot I_j(t)]},
	\end{equation}
	which can be checked by substituting all possible combinations of $A^0_{ij}$ and $\dot A_{ij}$.
	Divide by $\widetilde \Theta_i$ for both sides of Eq.~(\ref{qi_f}) and substitute $\widetilde \Theta_i=\prod_{j=1}^N(1-\lambda A_{ij} \widetilde I_j)$ gives
	\begin{eqnarray}\label{eq:pertuQ}
	1+\frac{\dot \Theta_i(t)}{\widetilde \Theta_i}&=&\prod_{j=1}^N\left(1-\frac{\lambda A_{ij}\dot I_j(t)}{1-\lambda A_{ij}\widetilde I_j}\right)  
	\times \prod_{j=1}^N\left(1-
	\frac{\lambda\dot{ A}_{ij}\dot I_j(t)}{1-\lambda\dot{ A}_{ij}\widetilde I_j}\right)^{-1} \nonumber \\
	& &\times \prod_{j=1}^N\left(1-\lambda\dot{ A}_{ij}\widetilde I_j\right)^{-1}.
	\end{eqnarray}
	Note that the following relation holds
	\begin{equation}\label{eq:AAA}
	\frac{\lambda A_{ij}\dot I_j(t)}{1-\lambda A_{ij}\widetilde I_j}=A_{ij}\frac{\lambda \dot I_j(t)}{1-\lambda \widetilde I_j},
	\end{equation}
	since $A_{ij}\in \{0,1\}$.
	Similarly, when replacing $A_{ij}$ in Eq.~(\ref{eq:AAA}) by $\dot{ A}_{ij}\in \{0,1\}$, we get
	\begin{equation}
	\frac{\lambda \dot{ A}_{ij}\dot I_j(t)}{1-\lambda\dot{ A}_{ij}\widetilde I_j}=\dot{ A}_{ij}\frac{\lambda \dot I_j(t)}{1-\lambda \widetilde I_j}.
	\end{equation}
	Taking the logarithm on both sides of Eq.~(\ref{eq:pertuQ}), expanding to the first orders of $\delta p_i(t)$, $\delta q_i(t)$, and applying the above relation can give
	\begin{equation}\label{eq:pertuQ2}
	\frac{\dot \Theta_i(t)}{\widetilde \Theta_i}=-\sum_{j=1}^N A_{ij}\frac{\lambda \dot I_j(t)}{1-\lambda \widetilde I_j}+\sum_{j=1}^N \dot{ A}_{ij}\frac{\lambda \dot I_j(t)}{1-\lambda \widetilde I_j}\\
	-\sum_{j=1}^N \log \left(1-\lambda\dot{ A}_{ij}\widetilde I_j\right).
	\end{equation}
	The terms in the last summation can be checked to satisfy
	\begin{equation}
	\log \left(1-\lambda\dot{ A}_{ij}\widetilde I_j\right)=\dot{ A}_{ij}\log \left(1-\lambda \widetilde I_j\right).
	\end{equation}
	With the above calculations, Eq.~(\ref{eq:pertuQ2}) can be written in the matrix form as 
	\begin{equation}\label{eq:dqExpress}
	\dot \Theta(t)=-\lambda \widetilde \Theta\circ \left(A - \dot{ A}\right) \Psi \dot{ A}(t)
	-\widetilde \Theta\circ \dot{ A}\log(1-\lambda \widetilde I),
	\end{equation}
	where $\log(1-\lambda \widetilde I)$ is the vector obtained by taking the logarithm in each entry of $1-\lambda \widetilde I$. And $\Psi$ is the $N\times N$ diagonal matrix  with entries
	\begin{equation}\label{eq:Zdef}
	\Psi_{ij} =\delta_{ij}\frac{1}{1-\lambda \widetilde I_j},
	\end{equation}
	where 
	\begin{equation}
	\delta_{ij}=\left\{\begin{array}{lc}1&i=j\\0&i\neq j\end{array}\right..
	\end{equation}
	Substituting Eq.~(\ref{eq:dqExpress}) back into Eq.~(\ref{eq:fixPertuEq1}) yields the following iteration formula for $\dot{ I}(t)$:
	\begin{eqnarray}
	\dot{ I}(t+1)&=&(\widetilde \Theta-\gamma)\dot{ A}(t)
	+(1-\widetilde I)\circ\lambda \widetilde \Theta\circ \left(A-\dot{ A}\right) \Psi \dot{ A}(t) \nonumber \\
	& &+(1-\widetilde I)\circ \widetilde \Theta\circ \dot{ A}\log(1-\lambda \widetilde I).
	\end{eqnarray}

	\section*{References}
	\bibliographystyle{iopart-num}
	\bibliography{xianjiajun}
	
\end{document}